\documentclass[prb,twocolumn,showpacs,preprintnumbers,amsmath,amssymb,superscriptaddress,aps,10pt]{revtex4-1}

\usepackage{graphicx}
\usepackage{dcolumn}
\usepackage{bm}
\usepackage{natbib}
\usepackage{amsmath}
\usepackage{color}
\usepackage{ulem}

\begin{document}

\title[SERS on CNTs]{Tailored nano-antennas for directional Raman studies of individual carbon nanotubes}

\author{Nicola Paradiso}\email{nicola.paradiso@physik.uni-regensburg.de}
\author{Fatemeh Yaghobian}
\author{Christoph Lange}
\author{Tobias Korn}
\author{Christian Sch\"uller}
\author{Rupert Huber}
\author{Christoph Strunk}
\affiliation{Institut f\"ur Experimentelle und Angewandte Physik, University of Regensburg}

\begin{abstract}	
\textit{We exploit the near field enhancement of nano-antennas to investigate the Raman spectra of otherwise not optically detectable carbon nanotubes (CNTs). We demonstrate that a top--down fabrication approach is particularly promising when applied to CNTs, owing to the sharp dependence of the scattered intensity on the angle between incident light polarization and CNT axis. In contrast to tip enhancement techniques, our method enables us to control the light polarization in the sample plane, locally amplifying and rotating the incident field and hence  optimizing the Raman signal. Such promising features are confirmed by numerical simulations presented here. 
The relative ease of fabrication and alignment makes this technique suitable for the realization of integrated devices that combine scanning probe, optical, and transport characterization.}
\end{abstract}

\pacs{78.67.Ch, 78.30.-j, 63.22.Gh, 61.48.De, 52.40.Fd, 42.79.Dj}

\maketitle

Similarly to antennas for communication systems, the operating principle of nano--optical devices consists in the ability to convey, focus, control, and re--emit electromagnetic signals.~\cite{NovotnyPT2011,NovotnyNP2011} Interestingly, it is often the near field of nano-antennas which is exploited in applications, as it can be used to greatly enhance the field amplitude and custom-tailor the electric field distribution locally. The size and shape of the antennas can be designed  to excite a plasmon resonance\cite{Murray2007} or to focus~\cite{Schnell2011} the optical near field to selected \textit{hot spots}.~\cite{Stockman1996}

A promising and direct application of nano-plasmonics is the surface-enhanced Raman spectroscopy (SERS),~\cite{Fleischmann1974,Moskovits1985,Stosch2011,FY2011} where metallic structures are used as antennas in order to enhance both the cross section and the field intensity~\cite{KneippRoyalSoc,CorioPRB2000} for Raman scattering.
The most effective structures employed in early SERS studies were colloidal gold or silver particles,~\cite{Abe1981,Kneipp1996,Kneipp2000} which provide dramatic enhancement factors in correspondence to randomly distributed hot spots. For many applications this approach appears insufficient: to fully unfold the potential of modern nano-optics technology, it is desirable to fabricate tailored devices that locally manipulate the optical field in correspondence to a specific target region of the sample.

Such a top-down approach to SERS is at present an emerging field.~\cite{Schedin2010,Heeg2013,Chen2013,Pors2014} First investigations employ patterned antenna structures to enhance the Raman signal of graphene.~\cite{Heeg2013} This particular choice is motivated by the fact that the Raman spectrum of graphene is well known, and it is relatively easy to transfer a single layer on top of selected patterned structures.
However, owing to their extremely anisotropic optical properties, carbon nanotubes (CNTs) appear in many respects to be more intriguing candidates for target structures.~\cite{Heeg2014,Hartmann2012,Mauser2014,Rai2012}
Raman spectroscopy of individual CNTs is usually difficult, due to their small cross section. The measured signal is significant only when either the incident or the scattered light is in resonance with an interband transition.~\cite{JorioPRL2001} Raman spectra are typically obtained from an ensemble of CNTs, from which one can occasionally detect the signal of an individual CNT which satisfies the resonance condition. 
For this reason CNTs highly benefit from the field amplification provided by tailored optical antennas,\cite{Heeg2014} that enable Raman signal enhancement of a \textit{given} CNT at a specific position, e.g.~next to selected electrodes. This opens new scenarios for experiments that combine optical and transport measurements. Local SERS can be used to determine diameter, chirality and number of defects for the portion of CNT contacted on a transport device. 
This information is often crucial for the interpretation of transport experiments,~\cite{Cao2004} since many important physical quantities (e.g.~spin-orbit interaction, Aharonov--Bohm flux, curvature-induced gap) depend on the diameter or on the chirality.~\cite{SaitoCNT,Laird2014}

Compared to tip-enhanced Raman spectroscopy (TERS),~\cite{Anderson2007,Hartschuh2003,Goss2011,Yano2006} the use of patterned nano-antennas introduces a new degree of freedom, i.e.~the possibility to rotate the electric field vector on the sub-wavelength scale. A mesoscopic field rotator could be harnessed to locally excite a specific CNT position. CNTs only absorb (and scatter) light with polarization parallel to their axis. Thus, by shining light polarized orthogonally to the axis, only the CNT portion in correspondence to the antennas will absorb light, while the rest of the CNT will effectively screen the radiation.

In the present work we investigate SERS of CNTs by making use of patterned arrays of nano-antennas. They allow us to acquire spectra even from CNTs whose Raman signal is too weak to be detected on the bare tube. From the amplified Raman spectra we can infer the local structure, the disorder and the metallicity of CNTs. By polarization-dependent Raman spectroscopy we demonstrate that antenna arrays effectively rotate the electric field vector of the incident light. The experimental data are quantitatively discussed in light of the results of numerical simulations.

\section{Experimental details}

\begin{figure*}[tb]
\includegraphics[width=2\columnwidth]{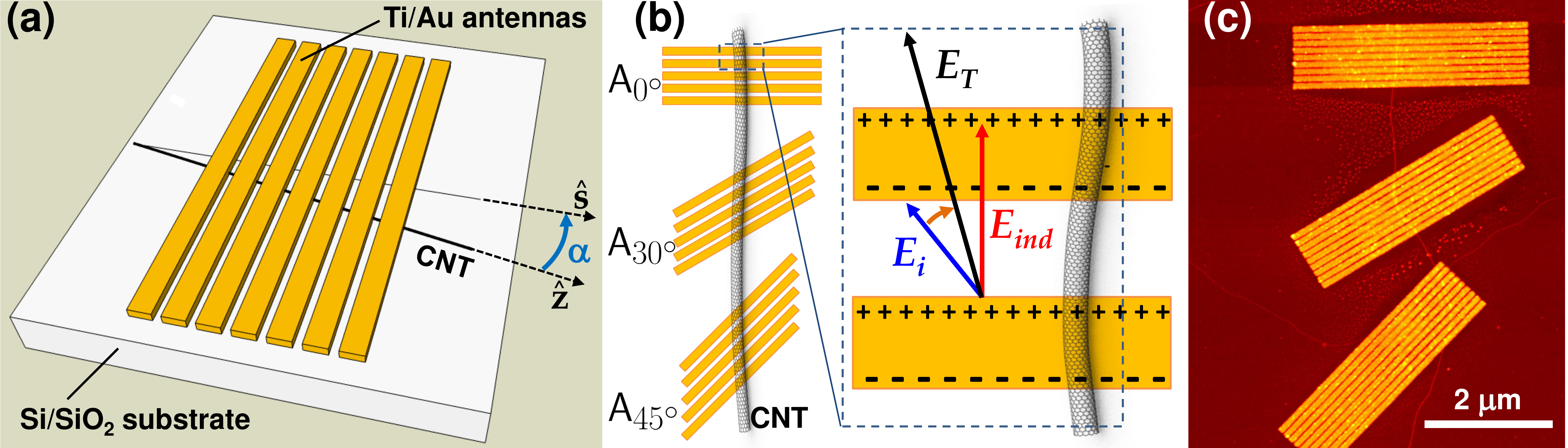}
\caption{(a) Sketch of a Ti/Au nano-antenna array patterned on the Si/SiO$_2$ substrate. The coupling with surface plasmon polaritons and thus the field enhancement depend on the angle $\alpha$ between the CNT axis $\mathbf{\hat{z}}$ and the direction $\mathbf{\hat{s}}$ orthogonal to the strips. (b) In some samples we have patterned three arrays $\mathsf{A}_{0^{\circ}}$, $\mathsf{A}_{30^{\circ}}$, and $\mathsf{A}_{45^{\circ}}$, whose angle $\alpha$ is 0$^{\circ}$, 30$^{\circ}$, and 45$^{\circ}$, respectively. Owing to the large aspect ratio of such structures, the incident  field $\mathbf{E_i}$ and the induced field $\mathbf{E_{ind}}$ are in general not parallel, since the latter is preferentially oriented along $\mathbf{\hat{s}}$.  (c) AFM micrograph of our sample $\mathcal{C}$, acquired after all measurements.~\cite{[{See the Appendix for further details}]SMnote1}}
\label{fig:sketch}
\end{figure*}

Figure~\ref{fig:sketch}(a) shows the geometry of our structures, consisting of arrays of metallic strips with a well-defined orientation with respect to the axis of a single-wall CNT.
The physical motivation behind choosing such a geometry is pictorially illustrated in Fig.~\ref{fig:sketch}(b). The incident light effectively polarizes the metal strips, whose thickness is of the order of the skin depth ($\approx 20$~nm for gold in the visible range). At optical frequencies, the strips can be seen as parallel plate capacitors that couple the far-field radiation into electromagnetic surface modes. Such correlated oscillations of charge density and electromagnetic field are called surface plasmon polaritons (SPPs).~\cite{Sobnack1998,Barnes2003} 
The periodic arrangement of antenna strips yields a spatially coherent standing plasmon wave which leads to a concentration of the electric field in the nanoscopic gaps between the strips. The resulting field enhancement within the array gaps is particularly useful when applied to Raman spectroscopy, since the Raman signal dependence on the field amplitude is non-linear. If both the incident and scattered signal are magnified by the same factor $\eta$, then the Raman signal intensity scales as $\eta^4$. 

The particular design of our antenna arrays allows us to control not only the field magnification, but also the polarization direction. The strip length is in fact much larger than both the optical wavelength and the spot size. We can therefore to first approximation neglect end-effects and consider the direction of the induced field as constant in the middle of the strips, and orthogonal to them, as shown in Fig.~\ref{fig:sketch}(b). Such local control of the light polarization is particularly interesting when applied to target structures which have an intrinsically anisotropic optical response, as in the case of CNTs. Owing to their large aspect ratio, CNTs can effectively screen only the field component orthogonal to the axis.  This property of CNTs (called \textit{depolarization effect}~\cite{Ajiki1995,Benedict1995}) makes them the ideal target to test the directionality of our antennas.

\begin{figure*}[tb]
\includegraphics[width=2\columnwidth]{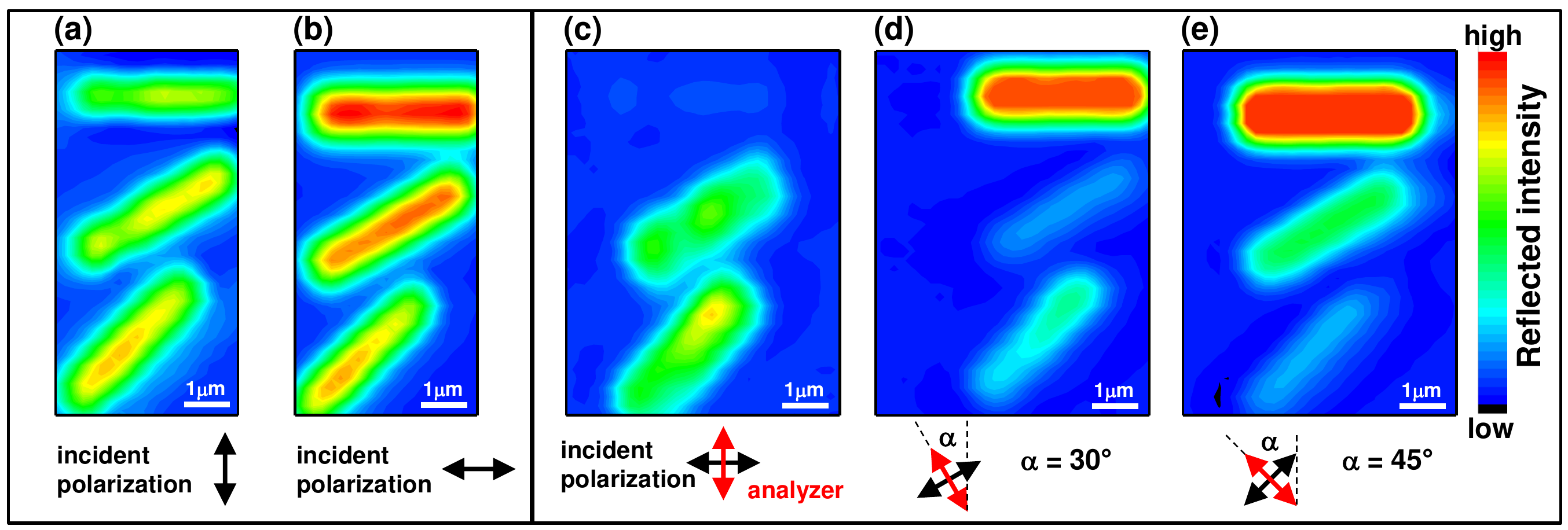}
\caption{ (a,b). Reflectance measurements performed on sample $\mathcal{C}$. The reflectance maps are obtained by scanning the sample while illuminating it with $\lambda_L=633$~nm light at normal incidence. The laser spot diameter is 1~$\mu$m. The panels (a) and (b) correspond to the reflected signal acquired for incident light polarization $\mathbf{\hat{e}_i}$ oriented orthogonally and parallel to the topmost antenna array, respectively. Both graphs show that the larger the angle between $\mathbf{\hat{e}_i}$ and $\mathbf{\hat{s}}$, the smaller is the SPP coupling. This implies a reduced absorption and thus a larger reflected signal. (c--e). Crossed-polarization reflectance measurements on the same sample, with $\mathbf{\hat{e}_i}$ forming an angle of $0^{\circ}$ (c), $30^{\circ}$ (d) and $45^{\circ}$ (e) with respect to the vertical direction. In this case the presence of an analyzer oriented orthogonally to $\mathbf{\hat{e}_i}$ allows us to observe only the signal with \textit{rotated} polarization. The reflectance maps demonstrate that the antennas effectively rotate the light polarization, except when this is nearly parallel to the strips.}
\label{fig:refl_meas}
\end{figure*}

Our samples are fabricated starting from a $p$-doped Si substrate with a 500~nm-thick SiO$_2$ cap layer. 
In a first lithographic step, the SiO$_2$ surface is patterned with an array of rhenium markers, which exhibit excellent
stability at the high temperatures required for the CNT growth. In a second lithographic step, we pattern an array of 1~$\mu$m-wide clusters of metallic nano-particles~\cite{[{See the Appendix for further details}]SMnote1} 
aligned to the rhenium markers. Afterwards, we grow the CNTs by chemical vapor deposition (CVD).~\cite{[{See the Appendix for further details}]SMnote1} This growth process yields almost exclusively single-wall CNTs, which are relatively straight and have a low number of defects compared to other techniques. For such reasons, the CVD growth became the standard method to synthesize CNTs for low-temperature transport measurements.~\cite{Laird2014} In our samples the average yield is about 1--2 long CNTs per catalyst dot, so that in one chip we typically have more than 100 tubes. CNTs start growing from the catalyst clusters and can extend up to tens of micrometers. Owing to van der Waals interactions, CNTs tend to stick to each other during growth and often form bundles of few CNTs in the vicinity of the catalyst dot, where their density is higher.

In order to align the antenna structures to CNTs, we perform initial AFM scans to locate and identify individual CNTs on the substrate. Although slower compared to scanning electron microscopy (SEM) imaging, the AFM characterization does not contaminate the sample and allows us to check the quality of the surface topography. In addition, AFM scans provide a coarse estimate of the CNT diameter and allow us to determine whether the CNTs are isolated or grouped in a bundle, since we are able to follow a given CNT from the catalyst particle to the end. 

Once a suitable CNT is found, nano-antenna structures are fabricated on top of it. An AFM scan of the final sample is shown in Fig.~\ref{fig:sketch}(c). The nano-antennas studied in the present work consist of gratings of about 90~nm--wide and 4~$\mu$m--long metal strips. The periodicity of the strips is 120~nm, so that the gap between them is approximately 30~nm. Each array contains 9 strips, which define a 1~$\mu$m wide and 4~$\mu$m long rectangular pattern. The strips are designed via electron beam lithography using the preliminary AFM scans as a reference to align each array to the CNT. The metal deposition is obtained by thermal evaporation of a Ti (4~nm)/Au (17~nm) bilayer. 

In most of our samples the antenna strips are oriented perpendicularly to the structures (as in the $\mathsf{A}_{0^{\circ}}$ array in Fig.~\ref{fig:sketch}(b)). For some particularly long CNTs (samples $\mathcal{A}$ and $\mathcal{C}$) we fabricated three arrays with different orientations. By defining $\mathbf{\hat{s}}$ as the unit vector perpendicular to the strips and $\mathbf{\hat{z}}$ as the CNT axis unit vector, the angle $\alpha$ between $\mathbf{\hat{s}}$ and $\mathbf{\hat{z}}$  is 0$^{\circ}$, 30$^{\circ}$, and 45$^{\circ}$ for the top ($\mathsf{A}_{0^{\circ}}$), middle ($\mathsf{A}_{30^{\circ}}$), and bottom ($\mathsf{A}_{45^{\circ}}$) array, respectively. The separation between the arrays has been designed to be larger than the laser spot size, in order to obtain distinct optical signals from different arrays or in between two of them.

Our optical setup is equipped with two laser sources with wavelengths $\lambda_L=532$ and 633~nm. The former is highly absorbed by gold structures, thus we only use the $\lambda_L=633$~nm wavelength for SERS measurements on antenna arrays.
The excitation beam is focused onto the sample through a 100$\times$ microscope objective, resulting in a spot size of about 1~$\mu$m. The polarization direction for the incident light is controlled via an achromatic $\lambda/2$ plate. Other details about our Raman setup have been reported elsewhere.~\cite{FY2012}
A given CNT with nano-antennas is easily found by means of Re markers. The determination of the CNT position is achieved by observing the antenna arrays with the objective and comparing the optical image with the preliminary AFM scans. With such a reference, we can place the laser spot exactly where the CNT crosses the patterned gratings, and acquire the corresponding local Raman or reflectance signal.

\section{Experimental results}

In order to prove the functionality of the nano--antennas, we have to demonstrate their ability to locally amplify the intensity and rotate the polarization of the incident light. 
The design of the antennas (and in particular the gap between the strips) has been chosen such that a SPP is excited when the array is illuminated with light at $\lambda_L=633$~nm. For the choice of the gap and strip width we refer to the work of Le Thi Ngoc \textit{et al.}~\cite{Ngoc2013} In this paper the authors present reflectance spectroscopy measurements on macroscopic arrays of gold strips. The experiment shows that an array of 80~nm-wide strips separated by a nominal gap of 20~nm is resonant at $\lambda_L \approx 630$~nm, i.e.~at such wavelength the observed SPP coupling is maximum. The authors also show that the resonating wavelength increases when the gap is decreased, while the resonance width decreases for larger strip width. We notice that for a nominal gap width of 20~nm, the actual one is roughly 30~nm,~\cite{Ngoc2013SM} similar to the actual gap of our metal strips. Therefore, we expect to observe in our antennas a strong SPP coupling around $\lambda_L=633$~nm. 
Furthermore, such coupling is expected to be anisotropic, owing to the elongated geometry of the metal strips. 

The emergence of SPP and its angular dependence can be visualized by performing local reflectance measurements, where the sample is illuminated with light ($\lambda_L=633$~nm) polarized along different angles. The 1~$\mu$m-wide laser spot is scanned over the antennas and the reflected intensity is measured by a photodiode for each position.
The results for sample $\mathcal{C}$ are shown in the panel (a) and (b) of Fig.~\ref{fig:refl_meas}. For vertical polarization (Fig.~\ref{fig:refl_meas}(a)), one expects to observe the best coupling for the $\mathsf{A}_{0^{\circ}}$ array, since in this case the induced charge polarization is orthogonal to the strip. The larger the angle $\gamma$ between $\mathbf{\hat{s}}$ and the incident polarization direction $\mathbf{\hat{e}_i}$, the smaller is the expected coupling with the SPP. A larger coupling implies a more pronounced absorption and thus less reflected signal. This is precisely what we observe: for incident light polarized vertically (Fig.~\ref{fig:refl_meas}(a)) the reflected signal decreases from the $\mathsf{A}_{45^{\circ}}$ array (lowest coupling) to the $\mathsf{A}_{0^{\circ}}$ array (optimal coupling). For horizontal polarization, the opposite is true (Fig.~\ref{fig:refl_meas}(b)).

As previously mentioned, the excitation of a SPP in our nano--antennas implies a rotation of the polarization of the scattered field, which is then mainly oriented perpendicularly to the strips. The polarization rotation can be directly imaged by crossed polarization reflectance measurements, where the reflected signal is filtered by an analyzer directed orthogonally to the $\mathbf{\hat{e}_i}$ vector. The results of such measurements are shown in the panels (c), (d), and (e) of Fig.~\ref{fig:refl_meas} for incident polarization directed orthogonally to the $\mathsf{A}_{0^{\circ}}$, $\mathsf{A}_{30^{\circ}}$ and $\mathsf{A}_{45^{\circ}}$ arrays, respectively. Clearly, the largest reflected signal is obtained for the antenna array directed at $45^{\circ}$ with respect to the incident field. The sharp angular dependence allows us to selectively suppress the signal coming from an individual array.

\begin{figure*}[tb]
\includegraphics[width=2\columnwidth]{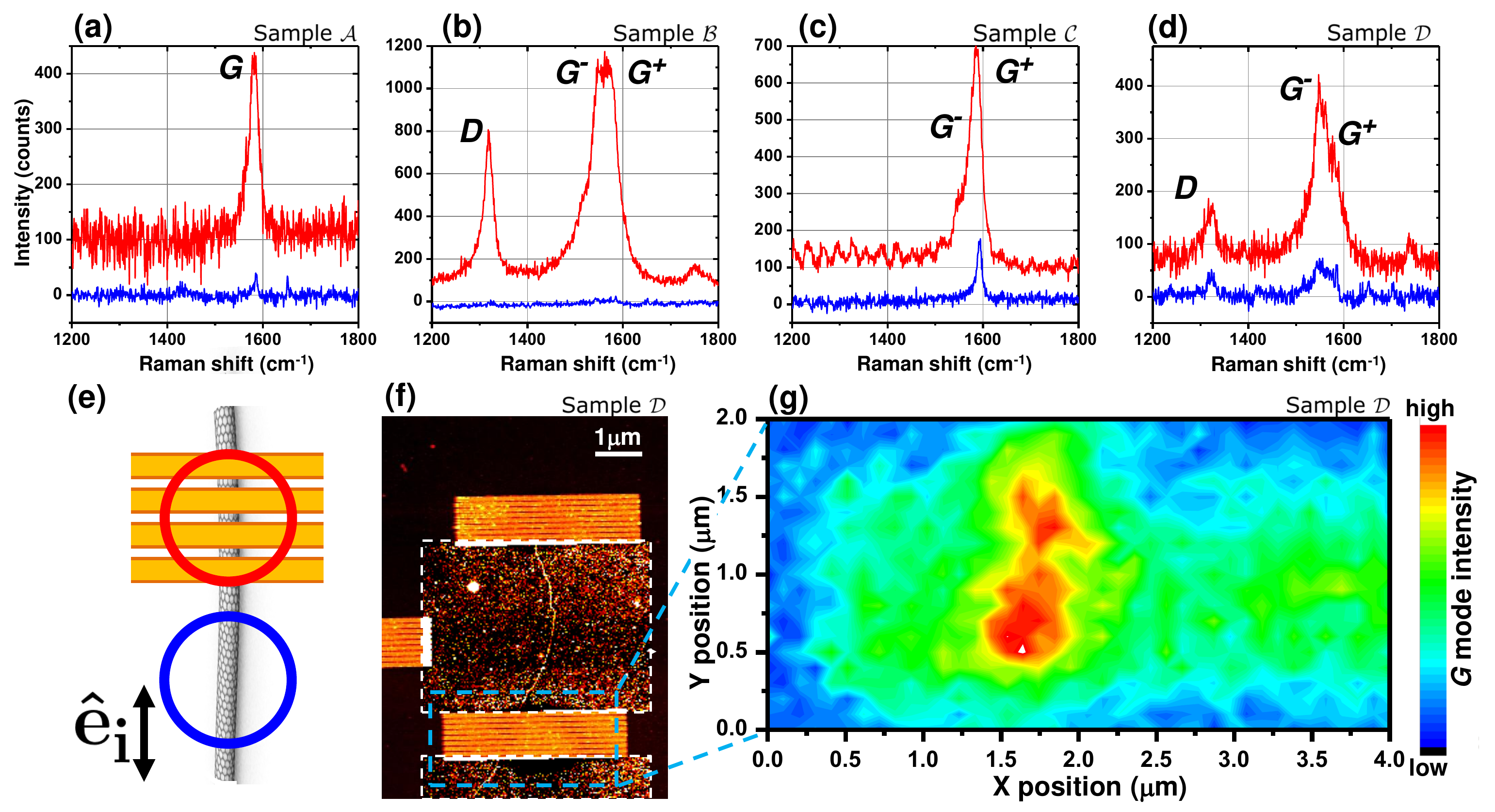}
\caption{Raman spectra measured on samples $\mathcal{A}$ (a), $\mathcal{B}$ (b), $\mathcal{C}$ (c), and $\mathcal{D}$ (d). The red curves correspond to the signal acquired on an antenna array orthogonal to the bundle, while the blue ones refer to the bare CNT portion immediately below the antennas, as sketched in panel (e).
For all the curves we subtract a background signal acquired next to the CNT. For clarity, a shift of $+100$ counts has been applied to all the enhanced curves. The integration time is 30~s for all the spectra. Panel (f) shows an AFM scan of sample $\mathcal{D}$. To make the CNT visible, we increase the contrast of the areas indicated by the dashed-white rectangles. For sample $\mathcal{D}$ we scan the laser spot over the antenna area highlighted (dashed-blue rectangle) in the AFM scan in panel (f). Panel (g) shows the corresponding color map for the intensity of the $G$ peak. A large signal is observed for the portion of CNT that crosses the antenna strips. The integration time is 5~s/pixel.}
\label{fig:SERS}
\end{figure*}

Reflectance measurements  indicate that SPPs are excited in the nano--antenna arrays. As a consequence, the optical field is considerably amplified within the antenna gaps. This has a dramatic impact when applied to SERS, due to the non-linear dependence of the scattered Raman signal on the field enhancement factor. 
In order to demonstrate the antenna-induced local enhancement, we perform a series of micro-Raman measurements on arrays fabricated on top of several individual CNTs or bundles. The results are summarized in Fig.~\ref{fig:SERS}. The panels (a), (b), (c) and (d) show Raman spectra for the $D$ and $G$ mode, acquired on the samples $\mathcal{A}$, $\mathcal{B}$, $\mathcal{C}$ and $\mathcal{D}$, respectively. From AFM scans we find that samples $\mathcal{A}$, $\mathcal{B}$ and $\mathcal{C}$ are most probably bundles of a few CNTs, whereas sample $\mathcal{D}$ is an individual CNT.~\cite{[{See the Appendix for further details}]SMnote1} The red curves in the reported spectra refer to the signal acquired on the portion of CNT or bundle underneath an array of orthogonal gold strips nominally identical to the $\mathsf{A}_{0^{\circ}}$ structure in Fig.~\ref{fig:sketch}(c).
The blue curves show the corresponding Raman spectrum acquired on a bare CNT portion just above or below the array, as sketched in Fig.~\ref{fig:SERS}(e). A crucial finding of our measurements is that the signal measured on the optical antennas is systematically larger than the one measured on the bare CNT, despite in the former case only one quarter of the CNT length (i.e.~the gap between two strips) is optically accessible.
We define the signal amplification factor $R$ as the ratio between the signal acquired on the antennas and the one measured on the bare CNT (or bundle).
For samples $\mathcal{A}$, $\mathcal{B}$, $\mathcal{C}$ and $\mathcal{D}$ we obtain $R=8$, 65, 3.5, and 5, respectively. The diversity of $R$ values will be discussed below in comparison to  the results of field simulations.

The antenna-induced amplification  allows us to extract useful information from samples whose spectrum would be otherwise barely discernible, as for the sample $\mathcal{A}$ and $\mathcal{B}$ in Fig.~\ref{fig:SERS}. A low intensity of the Raman signal indicates that the process is not (or not perfectly) resonant, i.e.~that the incident (or scattered) photon energy does not match any interband transition $E_{ii}$ of the CNT. For a given CNT and a given laser wavelength this is often the case. Optical antennas therefore allow one to increase the chances to observe the Raman spectrum from a specific CNT starting from a limited number of laser wavelengths.
In our experiment we acquire spectra at $\lambda_L=633$~nm from about 12 structures (individual CNTs or bundles of few CNTs) and we observe a clear Raman signal (i.e.~much larger than the background signal) for 7 structures, a weak signal for one structure and nearly no signal for 4 structures. By comparison, without antennas we observe a clear signal only for one structure (sample $\mathcal{C}$), a weak signal for 5 structures and no signal for all the other structures.

The measurement of the $D$ and the $G$ modes provides direct information about number of defects ($D$ mode) and geometry ($G$ mode) of CNTs. In particular, the $G$ mode is a fingerprint of the CNT chirality. Unlike graphene, the longitudinal optic (LO) and transverse optic (TO) component of the $G$ mode are no longer degenerate in CNTs. One of the two components (the TO for semiconducting CNTs, the LO for the metallic ones) corresponds to atom displacements along the circumference and is therefore softened by curvature effects.~\cite{Piscanec2007,Park2009,Sasaki2008} Therefore, instead of the single peak observed in graphene, in CNTs the $G$ mode is split in two components, $G^+$ and $G^-$, whose energy difference decreases with increasing the diameter $d$. In achiral (i.e.~zig-zag and armchair) CNTs, however, only one of the two modes is allowed by symmetry,~\cite{SaitoPRB1998} which makes them immediately identifiable in the spectra that show a single $G$ peak. In metallic CNTs, the mode at lower energy $G^-$ is considerably broadened and softened by the interactions with the electron Fermi sea.
Finally, the ratio between $G^+$ and $G^-$ has been predicted to depend on the chiral angle $\theta$, although there is no consensus yet about the precise functional form.~\cite{SaitoPRB2001,Telg2012,PailletStatSol2010}

From the SERS spectra in Fig.~\ref{fig:SERS} we can conclude that the bundle in sample $\mathcal{A}$ contains a semiconducting CNT which is relatively defect-free. Sample $\mathcal{B}$ contains a metallic CNT with a higher concentration of defects.
Sample $\mathcal{C}$ is of particular interest because the signal from the bare bundle is large enough to allow us to determine the impact of the antennas on the \textit{shape} of the $G$ mode. It is evident that the enhanced signal is much broader and displays more structure compared to the single sharp peak measured on the bare portion.~\cite{[{See the Appendix for further details}]SMnote1} In particular, the enhanced signal contains a metallic component (broad $G^-$ peak) which is absent in the bare bundle, whose spectrum corresponds to a semiconducting CNT with low chiral angle.~\cite{Park2009} The occurrence of a structural change in the chirality of a CNT can be ruled out because the signal on the bare portion of sample $\mathcal{C}$ has been acquired in between the arrays $\mathsf{A}_{0^{\circ}}$ and $\mathsf{A}_{30^{\circ}}$. The shape of the $G$ peak measured on the antennas above and below the bare portion is identical.~\cite{[{See the Appendix for further details}]SMnote1} We can therefore conclude that the antenna arrays enhance the Raman signal from different CNTs by different factors.

The interpretation of the enhanced spectra is more straightforward for an individual CNT, as in sample $\mathcal{D}$. In this case, the observed CNT is clearly a chiral metallic one with a moderate number of defects. The enhanced $G$ peak has roughly the same shape of the peak measured on the bare CNT.
On this sample, we also perform a Raman mapping of the $G$ mode by scanning the laser spot over the area indicated by the dashed rectangle in Fig.~\ref{fig:SERS}(f). The color map in Fig.~\ref{fig:SERS}(g) shows the values of the integrated intensity of the $G$ peak for each position of the laser spot. We notice that the signal is particularly intense only where the CNT intersects the antenna strips, whereas it is relatively low above and below the array.

The ultimate goal of Raman measurements on CNTs is the determination of the chiral indices $(n,m)$ which fully determine the chiral angle $\theta$ and diameter $d$. More than a decade ago, Jorio \textit{et al.}~\cite{JorioPRL2001} described a method to assign the chiral indices based on the radial breathing mode (RBM) only. 
This method can only be applied to CNTs with an interband transition very close to the incident photon energy. Since the resonance window where the RBM is observable is particularly narrow (approximately 60~meV),~\cite{MaultzschPRB2005} this condition is in general fulfilled only by a small fraction of a given set of CNTs. Several authors~\cite{MaultzschPRB2005,MichelPRB2009,PailletStatSol2010,ReichCNT} have shown that for individual CNTs the RBM alone is not sufficient to unambiguously determine the chiral indices. The information obtained from RBM must be combined with that deduced from $G$ mode features, i.e.~the splitting and the relative intensity of the $G^+$ and $G^-$ peaks, and the width and shape of the $G^-$. For instance, in our experiment the combination of RBM and $G$ mode data (measured at $\lambda_L=532$~nm) allows us to conclude~\cite{[{See the Appendix for further details}]SMnote1} that the bundle $\mathcal{B}$ likely contains a (22,15) CNT. The bundle $\mathcal{C}$ contains a semiconducting CNT with either (26,0) or (25,2) chiral indices ($\lambda_L=532$~nm) and a metallic CNT with either (21,6) or (20,8) chiral indices ($\lambda_L=633$~nm). For the last two cases, the pairs of possible assignments are adjacent members of the same family $p \equiv 2n+m$, where $p$ equals 52 and $48$ for the semiconducting and the metallic CNT, respectively. Adjacent members of the same family share very similar properties and are therefore hard to distinguish by Raman spectroscopy.~\cite{MaultzschPRB2005} Further details about our assignment procedure are discussed in the Appendix.

We stress that for many applications the knowledge of the exact chiral indices is not strictly required. What is often needed is the metallicity, the density of defects, the approximate diameter and chiral angle, and eventually the presence of other CNTs. This information can be deduced from the $D$ and $G$ mode profiles, for which the possibility of a signal magnification is useful in case of experimental setups with only a few laser lines.
On the other hand, the RBM signal acquired on the antennas is masked by a large background due to the light elastically scattered by the antenna surface. The tails of the large peak at zero Raman shift hinder the observation of low-frequency modes like the RBM. Thus, all the RBM peaks measured in our experiment have been acquired on the bare portions of the CNTs.

\begin{figure}[b!]
\includegraphics[width=0.95\columnwidth]{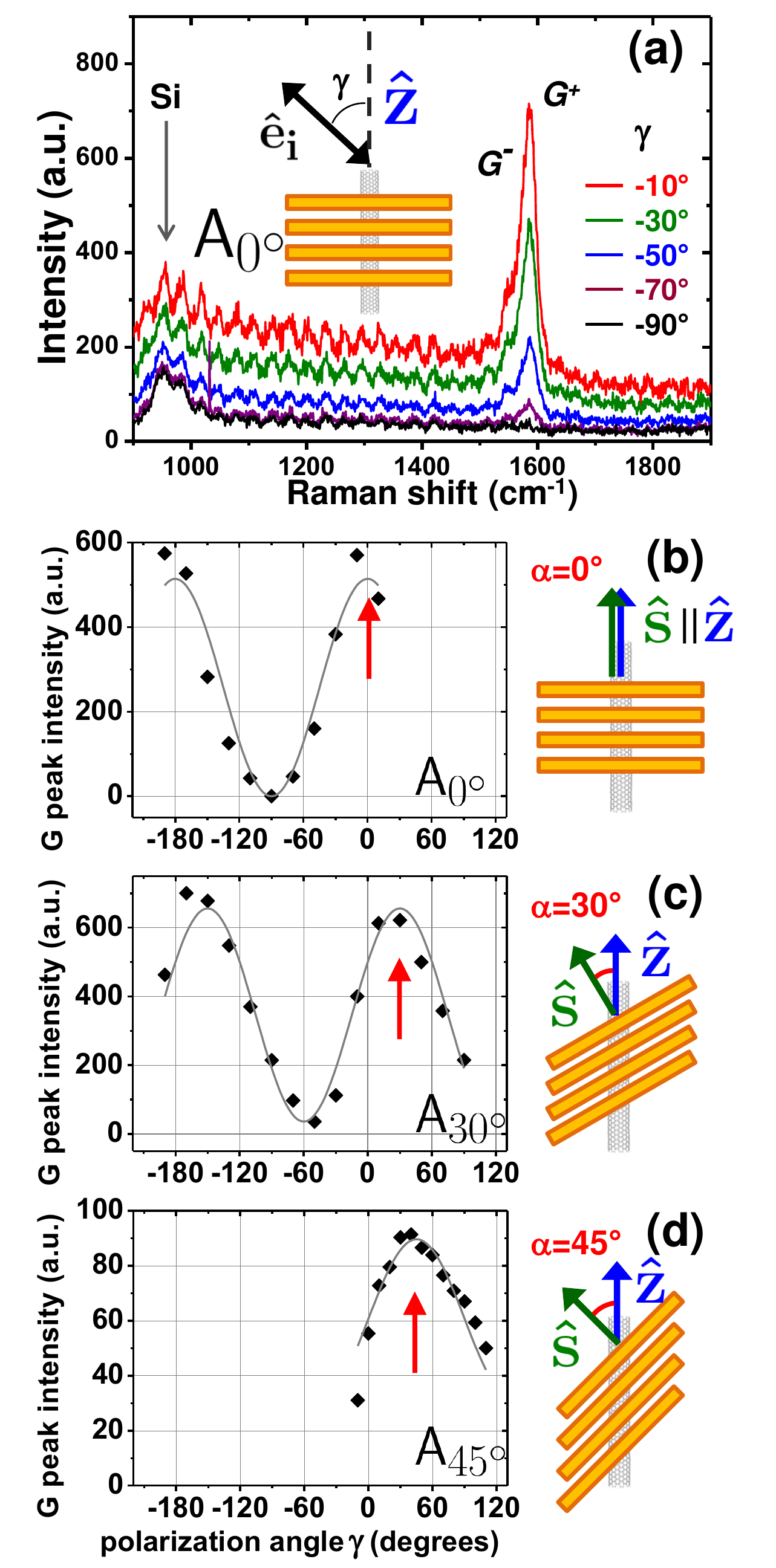}
\caption{(a) SERS spectra measured on the array $\mathsf{A}_{0^{\circ}}$ of sample $\mathcal{C}$ for different angles $\gamma$ between the CNT axis $\mathbf{\hat{z}}$ and incident polarization direction $\mathbf{\hat{e}_i}$. The small oscillations at lower Raman shifts are due to the signal elastically scattered by the antennas.   (b) $G$ peak amplitudes plotted as a function of $\gamma$ (black dots). The same measurement has been repeated for the array $\mathsf{A}_{30^{\circ}}$ (c) and $\mathsf{A}_{45^{\circ}}$ (d). The grey curves in panels (b), (c), (d) correspond to a $\cos^2(\gamma-\alpha)$ fit, with $\alpha=0^{\circ}$, $30^{\circ}$ and $45^{\circ}$, respectively. All data have been suitably rescaled to take into account the dependence of the incident power on the $\lambda/2$ plate rotation.~\cite{[{See the Appendix for further details}]SMnote1} Finally, the dark counts have been subtracted from all curves.}
\label{fig:pol_dep}
\end{figure}

The anisotropic optical response of CNTs makes them ideal target samples to study the impact on SERS of the antenna orientation  with respect to either the CNT axis  $\mathbf{\hat{z}}$ or the incident polarization direction $\mathbf{\hat{e}_i}$.  
Figure~\ref{fig:pol_dep} shows how the Raman signal measured on sample $\mathcal{C}$ depends on the angle $\gamma$ between  $\mathbf{\hat{e}_i}$ and $\mathbf{\hat{z}}$. Panel (a) displays selected Raman spectra measured on the $\mathsf{A}_{0^{\circ}}$ array for different $\gamma$ angles. The graph in Fig.~\ref{fig:pol_dep}(b) shows how the measured amplitude of the $G$ mode depends on $\gamma$.
The maximum Raman signal (expressed in terms of the $G$-peak amplitude) is measured for $\gamma \approx 0^{\circ}$ and $\gamma \approx -180^{\circ}$, while it is clearly suppressed for $\gamma \approx -90^{\circ}$. The experimental points can be fitted by a $\cos ^2\gamma$ function (grey curve in the graph). The $\cos^2\gamma$ dependence is expected for two reasons: on the one hand, the coupling between incident light and SPP is optimal for incident light polarized orthogonally to the array strips, while it is negligible for polarization parallel to the strips. On the other hand, owing to the depolarization effect, CNTs effectively screen the optical field directed orthogonally to their axis. As a consequence, the same angular dependence is observed also for the bare portion of the bundle.~\cite{[{See the Appendix for further details}]SMnote1} 

The situation changes for the $\mathsf{A}_{30^{\circ}}$ and $\mathsf{A}_{45^{\circ}}$ arrays, where the CNT axis $\mathbf{\hat{z}}$ is no longer parallel to the array axis $\mathbf{\hat{s}}$. Thus, for negligible field enhancements one would expect to observe the maximum Raman intensity for $\mathbf{\hat{e}_i} \parallel \mathbf{\hat{z}}$. On the other hand, for sizable enhancements the incident field $\mathbf{E_i}\equiv E_i\mathbf{\hat{e}_i}$ is negligible compared to the SPP-induced field ($\mathbf{E_{ind}}$, which is parallel to $\mathbf{\hat{s}}$) so that the total field is in this case parallel to the array axis $\mathbf{\hat{s}}$. The data in Fig.~\ref{fig:pol_dep}(c) and (d) are better described by the latter scenario: the $\cos^2\gamma$ behavior is reproduced, but the maximum intensity is observed at around $\gamma=\alpha$, i.e.~at $\gamma = 30^{\circ}$ and $\gamma = 45^{\circ}$ for the array $\mathsf{A}_{30^{\circ}}$ and $\mathsf{A}_{45^{\circ}}$, respectively.

\section{Discussion}
\begin{figure*}[tb]
\includegraphics[width=2\columnwidth]{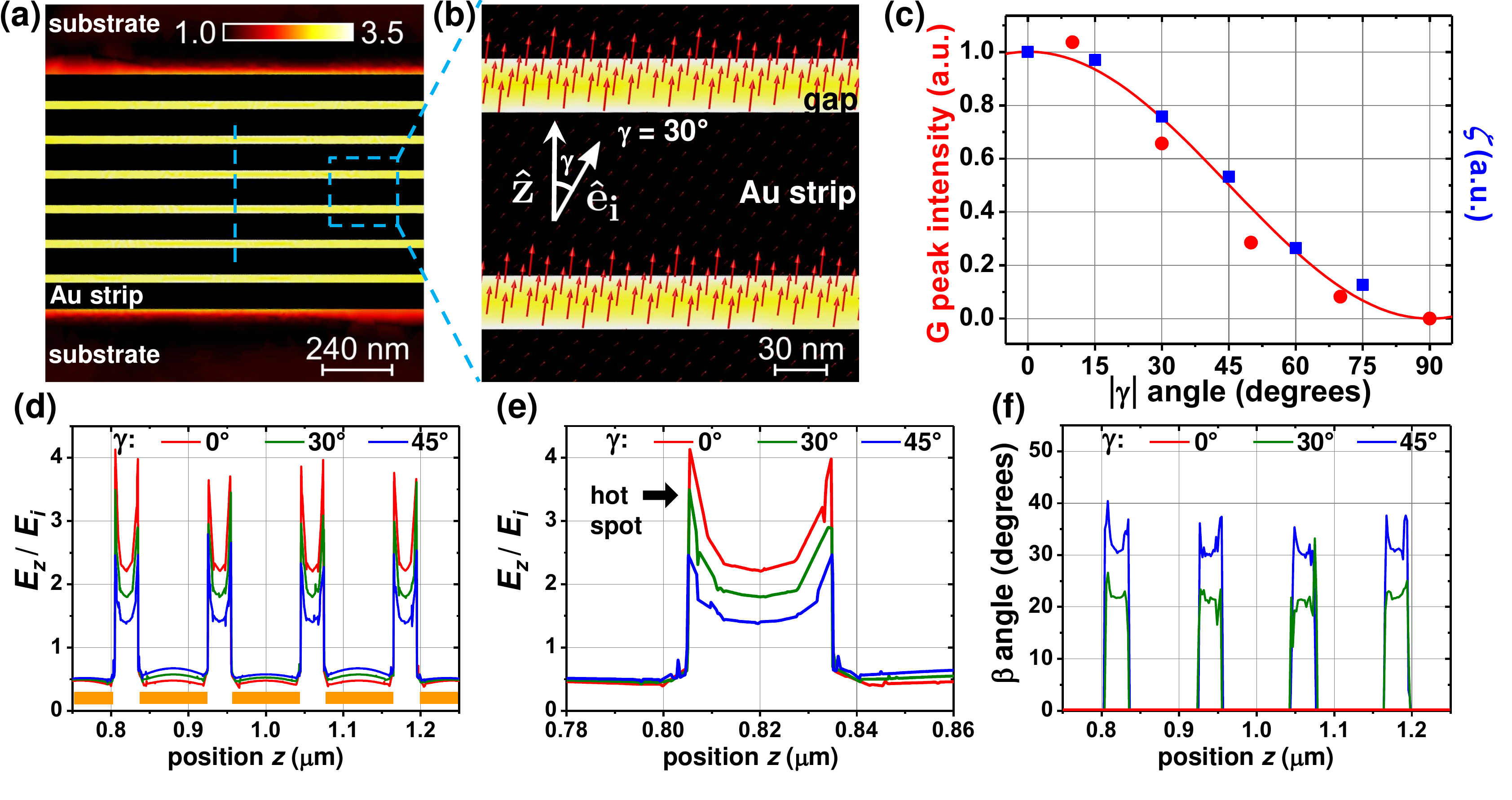}
\caption{ (a) Finite--difference frequency--domain calculation for the specific case of $\gamma=30^{\circ}$. The color plot shows the local value of $|\mathbf{E}_T|/E_i$, where $\mathbf{E}_T$ is the total electric field. (b) Zoom-in on the region indicated by the cyan-dotted square in panel (a). The arrow pattern shows the local orientation of the $\mathbf{E}_T$ vectors (red arrows). (c) $|\gamma|$ dependence (blue squares) of the Raman cross section for the array $\mathsf{A}_{0^{\circ}}$, calculated by integrating $(E_z(z)/E_i)^2$ over the cyan-dotted segment in panel (a). For comparison, we display (red dots) the experimental points (Fig.~\ref{fig:pol_dep}(b)) together with a $\cos^2\gamma$ fit (red curve) \textit{of the measured data}. 
The graph has been normalized such that the value for $\gamma=0^{\circ}$ equals 1 for both experimental and simulated data.
(d) Simulated $E_z(z)/E_i$ values plotted as a function of the $z$ coordinate along the segment highlighted in the panel (a). The red, green and blue curves correspond to the $\gamma=0^{\circ}$, $\gamma=30^{\circ}$ and $\gamma =45^{\circ}$ cases, respectively. (e) Zoom-in on a gap between two strips. The field is mostly concentrated within approximately 5~nm around the strip edge, in the so-called \textit{hot spots}.  (f) Calculated angle $\beta$ between $\mathbf{E}_T$ and $\mathbf{\hat{e}_i}$ plotted as a function of $z$ along the same range and for the same $\gamma$ values as in the panels (d,e).}
\label{fig:5sim}
\end{figure*}

In order to better understand the SERS mechanism in the antenna arrays and to interpret the amplitude ratios of our spectra, we perform finite-difference frequency-domain (FDFD) simulations to calculate the total field in the vicinity of an array of gold strips for several orientations of laser polarization $\mathbf{\hat{e}_i}$. 
The field enhancement is calculated for an incoming normalized plane wave polarized along the unit vector  $\mathbf{\hat{e}_i}$ forming an angle $\gamma$ with the $\mathbf{\hat{z}}$ direction. The antenna strips are assumed to be orthogonal to the CNT (i.e.~$\mathbf{\hat{s}} \parallel \mathbf{\hat{z}}$, as for the $\mathsf{A}_{0^{\circ}}$ array).
The color scale of Fig.~\ref{fig:5sim}(a) shows the modulus of the total electric field $\mathbf{E}_T\equiv E_i\mathbf{\hat{e}_i}+\mathbf{E_{ind}}$, for the particular case of $\mathbf{\hat{e}_i}$ directed at $30^{\circ}$ with respect to $\mathbf{\hat{s}}$.
The direction and magnitude of the total field is shown in Fig.~\ref{fig:5sim}(b) as arrows superimposed on the amplitude color map.
We repeat the calculation for several values of $\gamma$. For each of them we calculate ${E}_{z} \equiv (\mathbf{E}_{T}\cdot \mathbf{\hat{z}})$, i.e.~the projection of the total electric field on the CNT axis. This is the only field component that can induce a Raman signal on CNTs, owing to the depolarization effect.

In Fig.~\ref{fig:5sim}(d,e) we plot the calculated $E_z/E_i$ as a function of the CNT axis coordinate $z$ for $\gamma=0^{\circ}$, $30^{\circ}$ and $45^{\circ}$. The graph clearly shows that the local field enhancement is large close to the edge of the metal strips, where the spikes occur. Here the local field enhancement  is approximately 4. For largest field enhancements we expect to have the most pronounced field rotation induced by the antennas. In other words, if $|\mathbf{E_{ind}}|/E_i\gg 1$, then $\mathbf{E_{ind}}\approx \mathbf{E}_{T}$ and therefore the angle $\beta$ between $\mathbf{E}_{T}$ and $\mathbf{\hat{e}_i}$ will tend to $\gamma$. Figure~\ref{fig:5sim}(f) shows the profile of $\beta(z)$ along the CNT axis. We notice that the higher the enhancement factor, the more the  angle $\beta$ approaches $\gamma$. This is consistent with the observations on macroscopic arrays of metal nanowires.~\cite{Ngoc2013}

In order to compare the simulation results with the experimental points in Fig.~\ref{fig:pol_dep}(b) we compute the quantity $\zeta \equiv \int (E_z(z)/E_i)^2dz$, i.e.~the incident field intensity integrated over the four central metal strips of the array $\mathsf{A}_{0^{\circ}}$. The $\zeta$ value is proportional to the total Raman cross section and consequently it can be used to estimate the relative Raman intensities as a function of $\gamma$. In fact, the scattered Raman signal is already directed along $\mathbf{\hat{z}}$ and does not affect the $\gamma$-dependence.  The graph in Fig.~\ref{fig:5sim}(c) shows a comparison between the measured $G$ peak amplitudes (red dots) for the $\mathsf{A}_{0^{\circ}}$ array of sample $\mathcal{C}$ and the corresponding computed $\zeta$ (blue squares), both plotted as a function of $\gamma$. We also show a $\cos^2\gamma$ fit for the experimental points. Although the fit is computed for the experimental data, the curve matches nicely the computed $\zeta(\gamma)$ points. This indicates that the impact of the angle $\gamma$ consists in approximately rescaling the local enhancement by a factor $\cos(\gamma)$.

While the $\gamma$-dependence of the relative Raman signal is correctly reproduced by our calculations, the interpretation of the absolute enhancement factors requires more care. In fact, the experimental values for the amplification factor $R$ are quite diverse, ranging from 3.5 to 65. Such variability can be explained by the characteristics of each antenna array, e.g.~contamination from the fabrication process, or surface roughness due to an imperfect lift-off of the antennas. We use the results of the field simulation to establish whether or not the observed variability of the enhancement factors can be ascribed to imperfections in the antennas. The field simulation allows us to estimate the ideal enhancement factor in the absence of imperfections. We assume that both the incident and the scattered light are amplified in the same way by the antennas. For the $G$ mode the energy difference between incident and scattered light is $\omega_{G}\approx 0.2$~eV, comparable to the width of the SPP resonance.~\cite{Ngoc2013}
Thus, if the optical antennas are effective for both frequencies, the resulting enhancement will be proportional to the fourth power of the local electric field.~\cite{Moskovits1985,Heeg2013} In this approximation, the calculated amplification factor $R$ expected from the model is about 14.~\cite{[{See the Appendix for further details}]SMnote1} In most of the measured samples we observe an amplification factor lower than 14, which could be due to imperfections. A few samples (e.g.~sample $\mathcal{B}$), however, display considerably \textit{larger} $R$ values. Such unexpectedly large enhancements can be explained by the so-called \textit{chemical} SERS:~\cite{KneippRoyalSoc} besides enhancing the electromagnetic field, optical antennas can also directly exchange electrons with the CNTs. The resulting increase of the Raman cross section provides an additional amplification mechanism.
This effect is much more pronounced in metallic CNTs, as demonstrated by SERS experiments with colloidal Ag or Au clusters.~\cite{KneippRoyalSoc,CorioPRB2000}    
Chemical SERS can explain the change in the $G$ peak shape observed for some bundles. If the signal from metallic CNTs is disproportionately amplified, one can observe the appearance of a broad $G^{-}$ component not present in the spectrum of the bare CNT, as seen for instance in sample $\mathcal{C}$.

\section{Conclusions}
In conclusion, we have demonstrated SERS on individual CNTs and bundles using top-down patterned nano-antenna arrays. We obtained field amplification factors between 3.5 and 65, which allowed us to characterize CNTs whose Raman signal was not, or just barely, detectable without antennas. 
Owing to the particular geometry of our antennas, the induced field is rotated with respect to the polarization of the incident light, as verified by polarization-dependent SERS measurements.

The fabrication of patterned nano-antennas can be directly applied to transport devices to extract structural information in the vicinity of the CNT portion of interest. 
Furthermore, the ability to locally manipulate both modulus and direction of the optical field can be exploited to extract the local Raman signal of generic target molecules distributed on a substrate, e.g.~by employing crossed-polarized configuration to suppress the signal everywhere but at the antenna position.

We want to thank C.~B\"auml for the help with the measurements, D.~Steininger for the assistance with the growth of  the CNTs, and J.~Maultzsch for helpful discussions. The work was funded by the European Union within the STREP 'SE2ND', and by the Deutsche Forschungsgemeinschaft within SFB689, SFB631, and GRK1570.

\appendix

\section{Carbon nanotube growth}

Our fabrication process begins with  electron beam lithography (EBL) of 20~nm-thick Re markers, which are used as a reference both for the alignment of the subsequent lithographic steps, and for the global positioning of the laser beam during the Raman measurements.
In a second step we exploit again EBL to deposit clusters of catalyst nano-particles on the sample surface. We spin two layers of poly-methyl-methacrilate (PMMA), whose thickness is 320 and 70~nm, respectively. We design the catalyst spots by EBL, using the Re markers as a reference. Then we dip the sample in a methanol solution of bis\-(acetyl\-acetoneato)\-di\-oxo\-molybdenum(VI), aluminum oxide and ferric nitrate nonahydrate. We dry the solvent out and we remove the PMMA mask in acetone.

Then we insert the sample in a quartz tube positioned inside a furnace. The chemical vapor deposition (CVD) is performed in a steady flow of methane (10 sccm) and hydrogen (20 sccm) at 850~$^\circ$C for 20 minutes. This recipe produces a number of CNTs per catalyst dot. In general, however, only a few of them are longer than 1~$\mu$m. Each dot has on average 1 CNT longer than 10~$\mu$m.  

\begin{figure*}[tb]
\includegraphics[width=1.8\columnwidth]{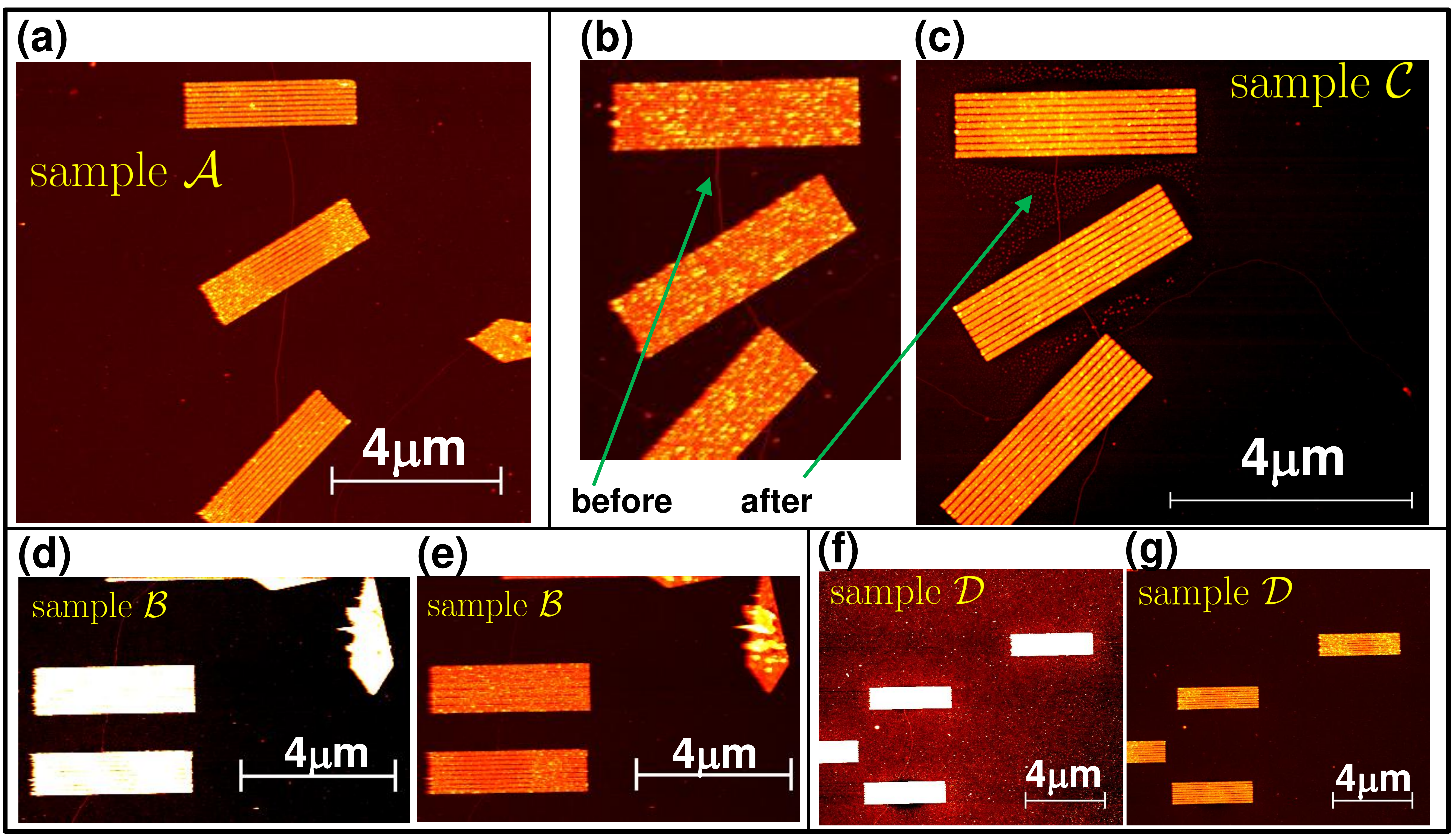}
\caption{Atomic force microscopy scans of the samples discussed in the present work. (a) AFM scan of sample $\mathcal{A}$ performed before the Raman measurements. (b) AFM scan of sample $\mathcal{C}$ performed before the Raman measurements. (c) The same sample scanned after all the measurement sessions. The topography reveals the presence of contamination exactly in the areas where the laser spot was focused. (d) AFM scan on sample $\mathcal{B}$. The contrast has been exaggerated in order to show the CNT. (e) The same scan displayed with lower contrast. (f) AFM scan on sample $\mathcal{D}$. The contrast has been exaggerated in order to show the CNT. (g) The same scan displayed with lower contrast.}
\label{fig:afm}
\end{figure*}

After the CNT growth we image the surface with an atomic force microscope (AFM) in tapping mode. These scans are used to check the presence of CNTs, their approximate diameter, and their exact position with respect to the Re markers. 
As last step, we fabricate the nano-antennas. We exploit the previous scans to correctly align the strip arrays on the Re markers (and therefore on the CNT). The antennas are fabricated by depositing 4~nm of Ti and 17~nm of Au by thermal evaporation.  
Owing to the small gap between the antenna strips, the lift-off of the metal layer is particularly critical. We perform a second AFM  scan in order to check the quality of these structures, and the actual CNT position within the gratings.

Figure~\ref{fig:afm} shows AFM topography scans of the samples studied in the present work. The panels (b) and (c) show the topography of sample $\mathcal{C}$ before and after the measurement session. We notice that contamination appears after the measurements which were not present \textit{before} (panel a). This contamination is localized exactly in the areas where the laser spot was focused for most of the time. Its origin is a sort of optical tweezer effect, i.e.~the attractive interaction that results from very intense optical fields on small particles in the vicinity.   This effect is relevant only for sample $\mathcal{C}$, which has been subjected to the long reflectance measurements. The Raman measurements, performed before the reflectance measurements, have not been affected by such contamination. 
From AFM analysis we deduce that: sample $\mathcal{A}$ has a diameter of ($5.0\pm0.5$)~nm; sample $\mathcal{B}$ has a diameter of ($2.5\pm0.5$)~nm; sample $\mathcal{C}$ has a diameter of ($4.5\pm0.5$)~nm; sample $\mathcal{D}$ has a diameter of ($1.5\pm0.3$)~nm.

\begin{figure}[tb]
\includegraphics[width=\columnwidth]{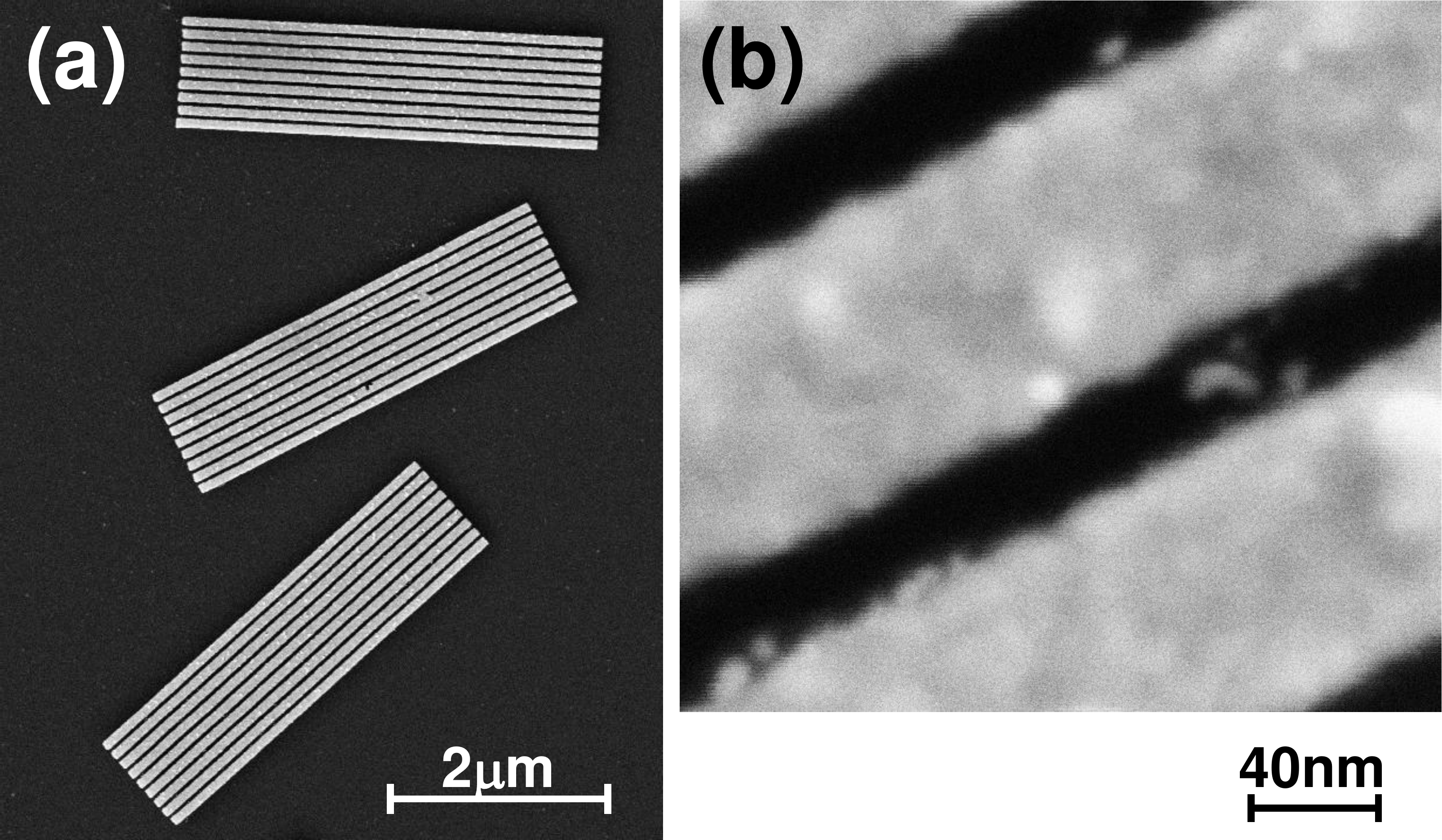}
\caption{(a) SEM micrograph of nano-antenna arrays nominally identical to the arrays in sample $\mathcal{C}$. (b) Zoom-in on the central array.}
\label{fig:sem}
\end{figure}

Figure~\ref{fig:sem} shows two scanning electron microscope (SEM) micrographs of a replica of sample $\mathcal{C}$, i.e.~a nominally identical device fabricated together with the original sample.

\section{Estimate of the chiral indices}

The assignment of the precise chiral indices to individual CNTs on a substrate is not a trivial task, in particular when there are only few laser lines available. More than a decade ago, Jorio \textit{et al.}~\cite{JorioPRL2001} proposed a method to assign the chiral indices based on the RBM only. The method relies on the fact that the Raman signal from individual CNTs is too weak to be detected, unless the incident laser light is in resonance with an interband transition of the CNT (resonant Raman scattering). The authors therefore start from a large ensemble of CNTs distributed on a substrate and perform micro-Raman measurements on several positions. If a clear RBM signal is measured in a specific position, the corresponding diameter is extracted. Then a Kataura plot is used to find which CNTs have an interband transition close (within 100~meV) to the laser photon energy. The information provided by the RBM measurements alone are certainly useful, but unfortunately its uncertainty is too large to allow one to safely determine the exact chiral indices. This has been pointed out several time in the literature.~\cite{MaultzschPRB2005,MichelPRB2009,PailletStatSol2010,ReichCNT}
For this reason, in recent years several experimental groups have developed alternative techniques (e.g.~Rayleigh scattering or electron diffraction) that provide and independent determination of the chiral indices for free-standing CNTs.

Nowadays it is clear that the determination of the chiral indices of individual CNTs by Raman spectroscopy must take into account all the available information: RBM frequency, $G$ peak shape, laser frequency. This data may be combined with AFM topography to yield accurate results. In the following paragraphs we explain the procedure we have followed to extract the chiral indices from samples $\mathcal{B}$ and $\mathcal{C}$.

\textbf{Sample $\mathcal{B}$, $\lambda_L=532$~nm}. AFM analysis shows that sample {$\mathcal{B}$} is a bundle with diameter of roughly 2.5~nm. As explained in the main text, the light elastically scattered from the antenna surface is particularly disturbing for low Raman shift, therefore the RBM measurements have been performed on the bare portion of the CNTs next to the antenna structure.
\begin{figure}[t!]
\includegraphics[width=\columnwidth]{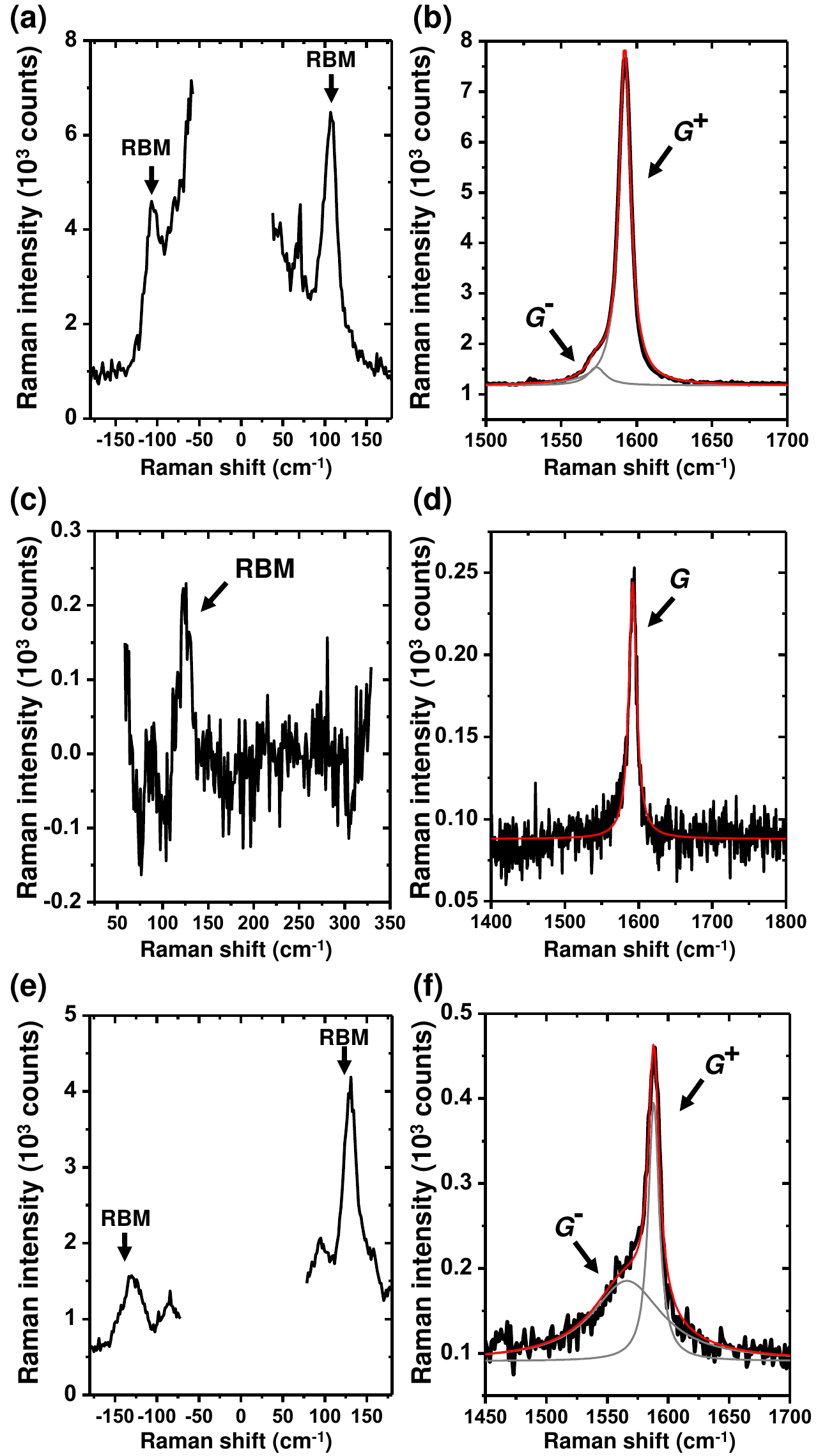}
\caption{(a) Raman spectrum measured on sample $\mathcal{B}$ using a laser source at $\lambda_L=532$~nm. The laser spot was placed on the bare portion of the bundle just below the antenna array. Both the Stokes and anti-Stokes peaks are visible. The RBM frequency is $\omega_{RBM}=(107\pm1)$~cm$^{-1}$. (b) Raman spectrum around the $G$ mode frequency, together with a double-peak lorentzian fit (red curve) showing the $G^-$ and $G^+$ components (grey curves). (c) Raman spectrum measured on the bare portion of sample $\mathcal{C}$ using a laser source at $\lambda_L=633$~nm. The RBM frequency is $\omega_{RBM}=(125\pm2)$~cm$^{-1}$. (d) Raman spectrum around the $G$ mode frequency, measured on the same position, together with a lorentzian fit (red curve). (e) Raman spectrum measured on the bare portion of sample $\mathcal{C}$ using a laser source at $\lambda_L=532$~nm. Both the Stokes and anti-Stokes peaks are visible. The RBM frequency is $\omega_{RBM}=(130\pm2)$~cm$^{-1}$. (f) Raman spectrum around the $G$ mode frequency, together with a double-peak lorentzian fit (red curve) showing the $G^-$ and $G^+$ components (grey curves).}
\label{fig:mergeI}
\end{figure} 
\begin{center}
	\begin{table}[tb]
  \begin{tabular}{|c|c|c|c|c|c|}
    \hline
    $(n,m)$ & $d$ & $\theta$ & [$E_{ii};i]$ & $E_L-E_{ii}$ & metallicity \\ \hline \hline
    (26,10) & 2.521~nm & 15.6$^{\circ}$  & [2.28~eV;5] & -0.050~eV & semicond. \\ \hline
    (22,15) & 2.525~nm & 23.8$^{\circ}$  & [2.34~eV;5] & 0.009~eV & semicond. \\ \hline
    (19,18) & 2.510~nm & 29.1$^{\circ}$  & [2.39~eV;5] & 0.059~eV & semicond. \\ \hline
    \hline   
  \end{tabular}
  \caption{Chiral indices for the CNTs within the window [$d=(2.52\pm0.04)$~nm, $E_{ii}=(2.33\pm0.100)$~eV]. The values of the interband transitions have been taken from Ref.~\cite{Liu2012}, with a 40~meV redshift correction to account for the substrate interaction.}
  \label{table:tableI}
  \end{table}  
\end{center}
In the case of sample $\mathcal{B}$, we focus the laser spot on the bundle segment just below the array. The measurement with red light did not reveal any clear RBM peak, while using green light ($\lambda_L=532$~nm) we find a peak at $(107\pm 1)$~cm$^{-1}$, as shown in Fig.~\ref{fig:mergeI}(a). The first problem is to convert this frequency into an estimate for the diameter. From the literature we know that the RBM frequency $\omega_{RBM}$ depends linearly on the reciprocal of the diameter $1/d$. However, the interaction with the substrate causes a slight deviation from the perfect proportionality.  A more accurate functional dependence is the following:~\cite{Saitobook} 
\begin{equation}
\omega_{RBM}= \frac{227}{d}\sqrt{1+Cd^2},
\label{eq:EqSaito}
\end{equation}
where $C$ is a substrate-dependent constant, $\omega_{RBM}$ is expressed in cm$^{-1}$ and the diameter in nm. The constant $C$ for SiO$_{2}$ reported by Saito \textit{et al.}~\cite{Saitobook} is 0.065. Such constant is however based only on a single experimental point.~\cite{JorioPRL2001} More recently Paillet \textit{et al.}~\cite{PailletPRL2006}  have made an extensive RBM study on many free-standing CNTs, using electron diffraction to independently determine the precise diameter. The empirical dependence found by Paillet \textit{et al.} is 

\begin{equation}
\omega_{RBM}= \frac{204}{d}+27,
\label{eq:EqPaillet}
\end{equation}
where again $\omega_{RBM}$ is expressed in cm$^{-1}$ and the diameter in nm. The two functions in Eq.~\ref{eq:EqSaito} and~\ref{eq:EqPaillet} are very similar in the diameter range of our interest (1--2~nm): their difference ($\approx$40~pm) is of the order of the uncertainty due to the typical error in the determination of the RBM peak center ($\approx$1--2~cm$^{-1}$).

Since for sample $\mathcal{B}$ we measure $\omega_{RBM}=(107\pm1)$~cm$^{-1}$, we deduce a diameter $d=(2.52\pm0.04)$~nm. In this diameter range we search for all the CNTs that have an interband transition $E_{ii}$ close to the laser photon energy $E_L=2.33$~eV ($\lambda_L=532$~nm). Following the literature,\cite{JorioPRL2001,MaultzschPRB2005} the window of acceptable values for $E_{ii}$ is set to $E_L\pm100$~meV, where 100~meV corresponds roughly to twice the width of the Raman resonance for the RBM.
To this end, we use the experimental interband values reported in recent extensive Rayleigh spectroscopy experiments.~\cite{Liu2012} The values have been corrected (i.e.~$\approx$40~meV red-shifted) in order to take into account the effect of the substrate.~\cite{[{K.~Liu, F.~Wang, private communication}]SMNote2}
Among all the possible candidates, we choose only semiconducting CNTs, since the $G$ peak profile measured with the same laser wavelength is compatible with a semiconducting CNT. In fact, the lorentzian fit in Fig.~\ref{fig:mergeI}(b) shows that the $G^-$ component occurs at a frequency well above 1550~cm$^{-1}$.  Table~\ref{table:tableI} lists the three CNTs which have diameter and an interband transition within the region of interest (i.e.~$d=(2.52\pm0.04)$~nm, $E_{ii}=(2.33\pm0.100)$~eV). Two of them [(26,10) and (19,18)] are however quite off-resonance ($|E_{ii}-E_L|>50$~meV), which does not seem compatible with the fact that we observe a RBM signal also with relatively low integration time ($<5$~s). Therefore we assign to this CNT the chiral indices (22,15). We stress that such chiral indices do not refer to the metallic CNT that produces the Raman spectrum displayed in Fig.~\ref{fig:refl_meas}. That spectrum was in fact acquired using the $\lambda_L=633$~nm laser source and therefore refers to another CNT in the same bundle.
Finally, we notice that the diameter of the (22,15) CNT ($d=2.52$~nm) is compatible with the bundle diameter found by AFM ($d\approx 2.5$~nm). This indicates that the (22,15) is the largest CNT of the bundle.

\textbf{Sample $\mathcal{C}$, $\lambda_L=633$~nm}. The sample $\mathcal{C}$ displays a RBM peak for both laser wavelengths at our disposal, i.e.~$\lambda_L=633$ and $532$~nm. 
\begin{center}
	\begin{table}[tb]
  \begin{tabular}{|c|c|c|c|c|c|}
    \hline
    $(n,m)$ & $d$ & $\theta$ & [$E_{ii};i]$ & $E_L-E_{ii}$ & metallicity \\ \hline \hline
    (26,0) & 2.036~nm & 0$^{\circ}$  & [1.98~eV;4] & 0.21~eV & semicond. \\ \hline
    (25,2) & 2.041~nm & 3.8$^{\circ}$  & [1.98~eV;4] & 0.021~eV & semicond. \\ \hline
    (24,4) & 2.055~nm & 7.6$^{\circ}$  & [1.98~eV;4] & 0.021~eV & semicond. \\ \hline
    (19,11) & 2.059~nm & 21.2$^{\circ}$  & [2.05~eV;4] & 0.091~eV & semicond. \\ \hline
    (23,6) & 2.077~nm & 11.3$^{\circ}$  & [2.00~eV;4] & 0.041~eV & semicond. \\ \hline
    (22,8) & 2.108~nm & 14.9$^{\circ}$  & [1.98~eV;4] & 0.021~eV & semicond. \\ \hline
    (18,13) & 2.112~nm & 24.7$^{\circ}$  & [2.04~eV;4] & 0.081~eV & semicond. \\ \hline
    \hline   
  \end{tabular}
    \caption{Chiral indices for the CNTs within the window [$d=(2.05\pm0.04)$~nm, $E_{ii}=(1.959\pm0.100)$~eV]. The values of the interband transitions have been taken from Ref.~\cite{Liu2012}, with a 40~meV redshift correction to account for the substrate interaction.}
    \label{table:tableII}
  \end{table}  
\end{center}
We start our analysis from the former case. The Raman spectrum around the RBM and the $G$ mode is displayed in Fig.~\ref{fig:mergeI}(c). The $G$ peak shape displays only one component: the $G^{-}$ component is either very low or absent. The Raman spectrum is well fitted by a single lorentzian peak, as shown in Fig.~\ref{fig:mergeI}(d). This corresponds to either an achiral CNT, or to a CNT with low chiral angle. This considerably restricts the number of possible candidates. The RBM peak is centered at $125$~cm$^{-1}$. The peak is weaker than in the previous case, thus we consider here a larger uncertainty, i.e.~$\pm2$~cm$^{-1}$. Using Eq.~\ref{eq:EqSaito}, we deduce a diameter $d=(2.05\pm0.04)$~nm. In the window [$E_{ii}=(1.959\pm0.100)$~eV; $d=(2.05\pm0.04)$~nm] we find 7 possible CNT candidates, listed in Table~\ref{table:tableII}. They are all semiconducting. Only two of them have a small chiral angle, compatible with the observation of a negligible $G^{-}$ component,~\cite{Park2009,SaitoPRB2001,Telg2012} namely (26,0) and (25,2). These are two adjacent elements of the same family $2n+m=52$, which share very similar properties and are therefore difficult to discriminate.  Therefore we consider both (25,2) and (26,0) as possible chiral indices for the spectra of Fig.~\ref{fig:mergeI}(c) and (d).

\textbf{Sample $\mathcal{C}$, $\lambda_L=532$~nm}. Using the laser source at $\lambda_L=532$~nm we obtain the Raman spectra in Fig.~\ref{fig:mergeI}(e) and (f) for the RBM and the $G$ mode, respectively. 
\begin{center}
	\begin{table}[tb]
  \begin{tabular}{|c|c|c|c|c|c|}
    \hline
    $(n,m)$ & $d$ & $\theta$ & [$E_{ii};i]$ & $E_L-E_{ii}$ & metallicity \\ \hline \hline
    (21,6) & 1.923~nm & 12.2$^{\circ}$  & [2.39~eV;2] & 0.059~eV & metallic \\ \hline
    (20,8) & 1.957~nm & 16.1$^{\circ}$  & [2.40~eV;2] & 0.069~eV & metallic \\ \hline
        \hline   
  \end{tabular}
  \caption{Chiral indices for the CNTs within the window [$d=(1.95\pm0.04)$~nm, $E_{ii}=(2.33\pm0.100)$~eV]. The values of the interband transitions have been taken from Ref.~\cite{Liu2012}, with a 40~meV redshift correction to account for the substrate interaction.}
    \label{table:tableIII}
  \end{table}
\end{center}
We notice that the $G$ mode is clearly metallic, owing to the presence of a broad $G^-$ component at around 1550~cm$^{-1}$, as shown in Fig.~\ref{fig:mergeI}(f). The RBM peak is centered at $(130\pm2)$~cm$^{-1}$, which corresponds to a diameter $d=(1.95\pm0.04)$~nm. Table~\ref{table:tableIII} shows the 2 metallic CNTs [(21,6) and (20,8)] within the window [$E_{ii}=(2.330\pm0.100)$~eV; $d=(1.95\pm0.04)$~nm]. Between them, the most likely assignment is (21,6), owing to the smaller $E_{ii}-E_L$ value.~\cite{[{We also considered the chiralities (25,1), and (24,2), which have a diameter close to the edge of our acceptance window ($d_{25,1}=1.999$~nm, $d_{24,2}=2.008$~nm) and $E_{ii}-E_L\approx0.04$~eV. However, they have a small chiral angle ($\theta_{25,1}=2.0^{\circ}$, $\theta_{24,2}=5.8^{\circ}$), which is not compatible with the sizable $G^{-}$ component observed}]noteotherCNT}

\section{Raman measurements on the bare portion of sample $\mathcal{C}$}
\begin{figure}[bt]
\includegraphics[width=1\columnwidth]{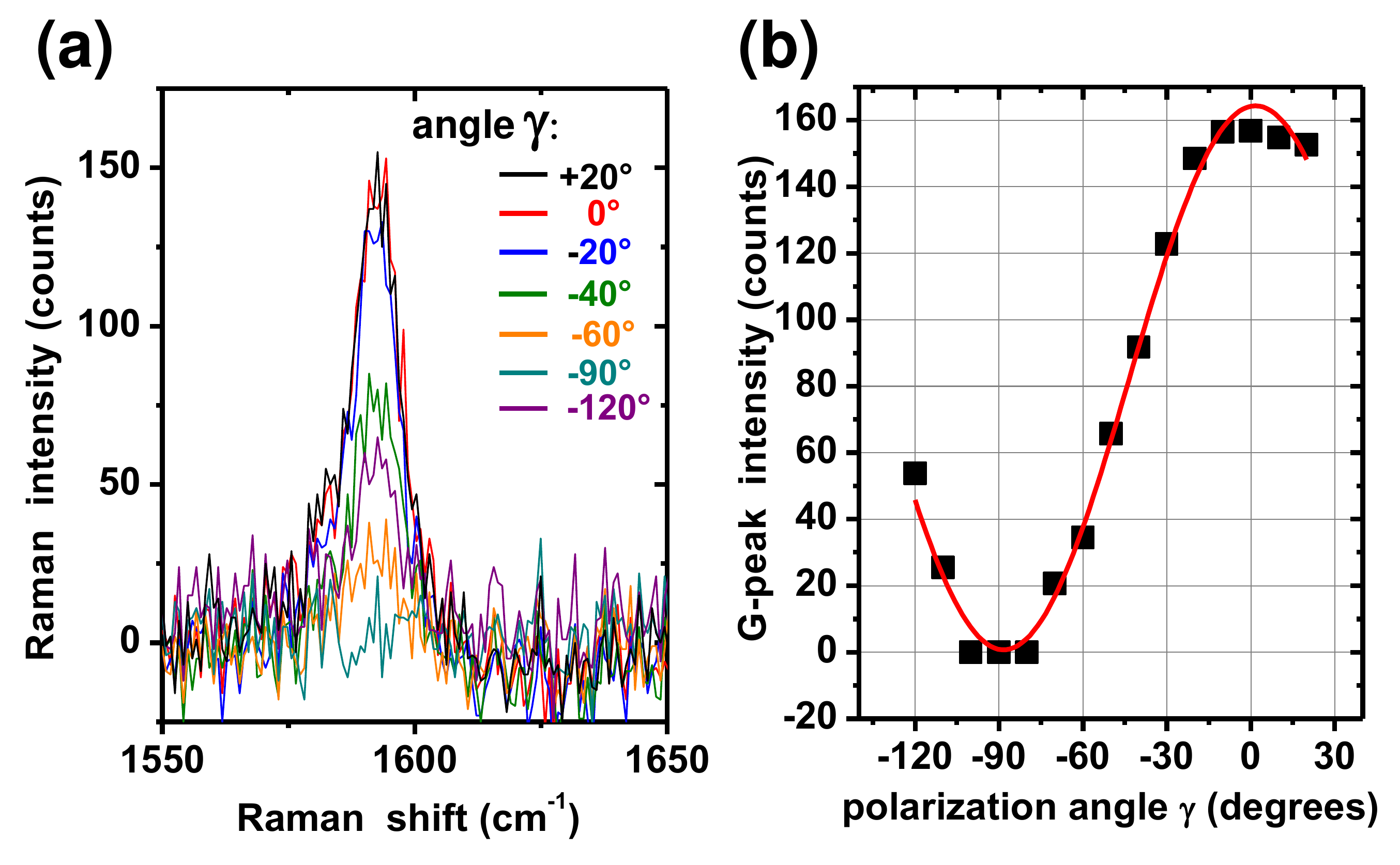}
\caption{(a) Raman spectra measured on the bare portion of sample $\mathcal{C}$ between the antenna array $\mathsf{A}_{0^{\circ}}$ and $\mathsf{A}_{30^{\circ}}$. Each curve refers to a different $\gamma$ angle. (b) $G$ peak amplitudes plotted as a function of $\gamma$ together with the best fit for $A\cos^2(\gamma+\phi)+I_0$, where $A$, $\phi$ and $I_0$ are fitting parameters (best fit: $A=164$, $\phi=1.7^{\circ}$, $I_0=0.7$).}
\label{fig:bare_pol_dep}
\end{figure}
Figure~\ref{fig:bare_pol_dep} shows the $\gamma$-dependence of the Raman spectrum around the $G$ mode for the portion of bare bundle in between the antenna arrays $\mathsf{A}_{0^{\circ}}$ and $\mathsf{A}_{30^{\circ}}$ in sample $\mathcal{C}$. The measurement completes the data shown in Fig.~\ref{fig:pol_dep}. Owing to the depolarization effect (discussed in the main text), the maximum signal is observed for $|\gamma|=0^{\circ}$, while for $|\gamma|=90^{\circ}$ it is completely suppressed. As expected the $\gamma$-dependence is well fitted by a $\cos^2\gamma$ function.

\begin{figure}[tb]
\includegraphics[width=1\columnwidth]{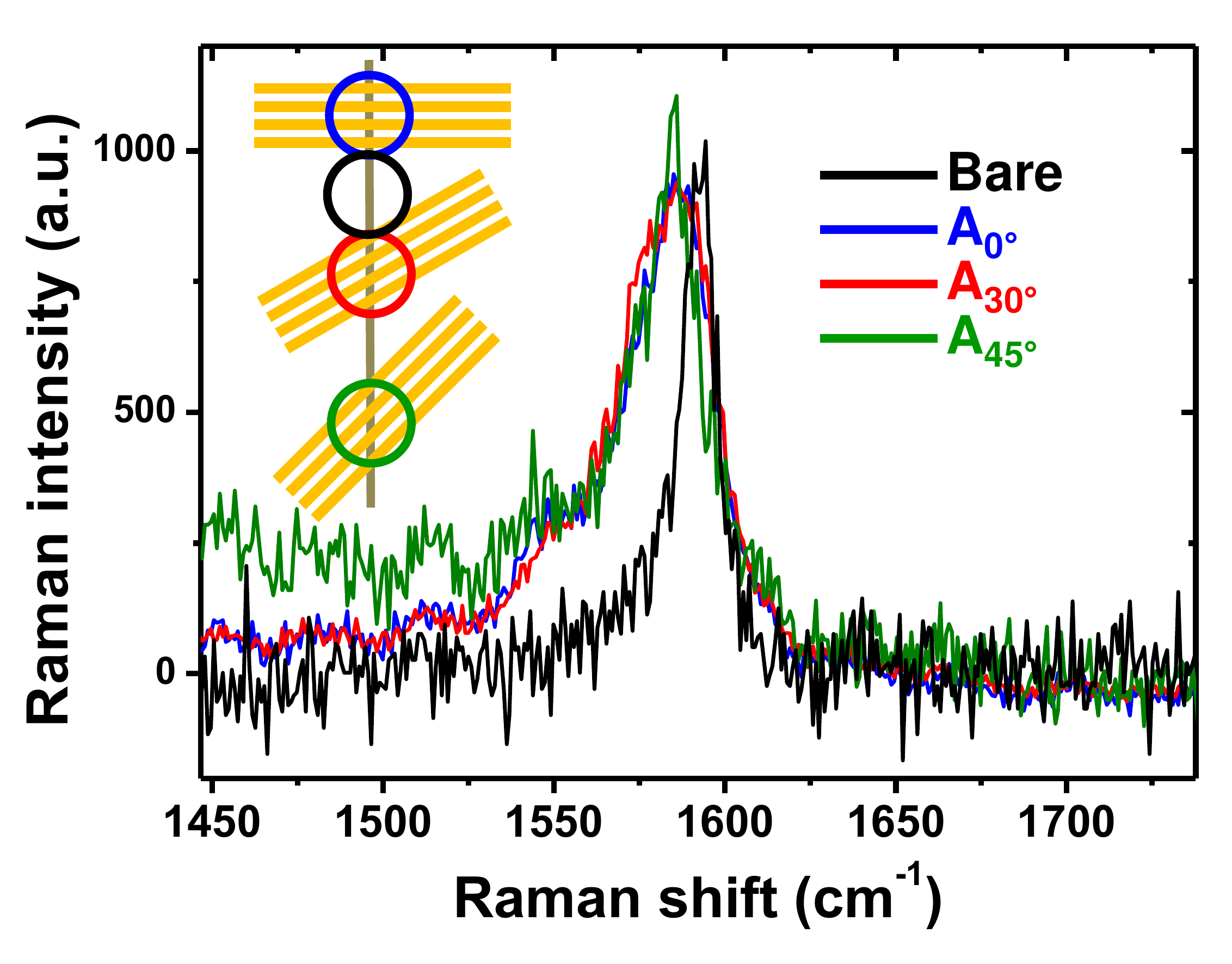}
\caption{Raman spectra around the $G$ peak measured on sample $\mathcal{C}$. The blue, red, green and black curve refer to the signal measured on $\mathsf{A}_{0^{\circ}}$, $\mathsf{A}_{30^{\circ}}$, $\mathsf{A}_{45^{\circ}}$ and on the bundle portion between $\mathsf{A}_{0^{\circ}}$ and $\mathsf{A}_{30^{\circ}}$, respectively. The curves have been offset and rescaled in order to display the same $G$ peak amplitude.}
\label{fig:comparison}
\end{figure}
Figure~\ref{fig:comparison} shows a comparison between a Raman spectrum measured on the $\mathsf{A}_{0^{\circ}}$ array of sample $\mathcal{C}$ (blue curve), the one acquired on the  $\mathsf{A}_{30^{\circ}}$ array (red curve), the one acquired on the  $\mathsf{A}_{45^{\circ}}$ array (green curve), and the one measured in the bundle portion between  $\mathsf{A}_{0^{\circ}}$ and  $\mathsf{A}_{30^{\circ}}$ (black curve). For better visibility we normalize the four curves to the same $G$ peak amplitude. We also subtract the background signal from all the curves, by measuring the signal next to the CNT. The graph shows that the spectra measured on the antennas are similar to each other, but they look quite different from the spectrum measured on the bare bundle. In particular the spectra measured on the antennas show a broad $G^-$ component at around 1550~cm$^{-1}$, a clear indication of a metallic CNT. On the contrary, the spectrum measured on the bare portion displays only one component. Since this portion is located in between the $\mathsf{A}_{0^{\circ}}$ and $\mathsf{A}_{30^{\circ}}$ arrays, we cannot explain this change as a structural change occurring to the same CNT, since the spectra on $\mathsf{A}_{0^{\circ}}$ and $\mathsf{A}_{30^{\circ}}$ are the same. In the main text we explain such change in terms of a disproportional SERS enhancement of a metallic CNT within the bundle due to the so-called \textit{chemical} SERS.

\section{SERS enhancement dependence on the  angle $\alpha$}
\begin{figure}[b]
\includegraphics[width=0.7\columnwidth]{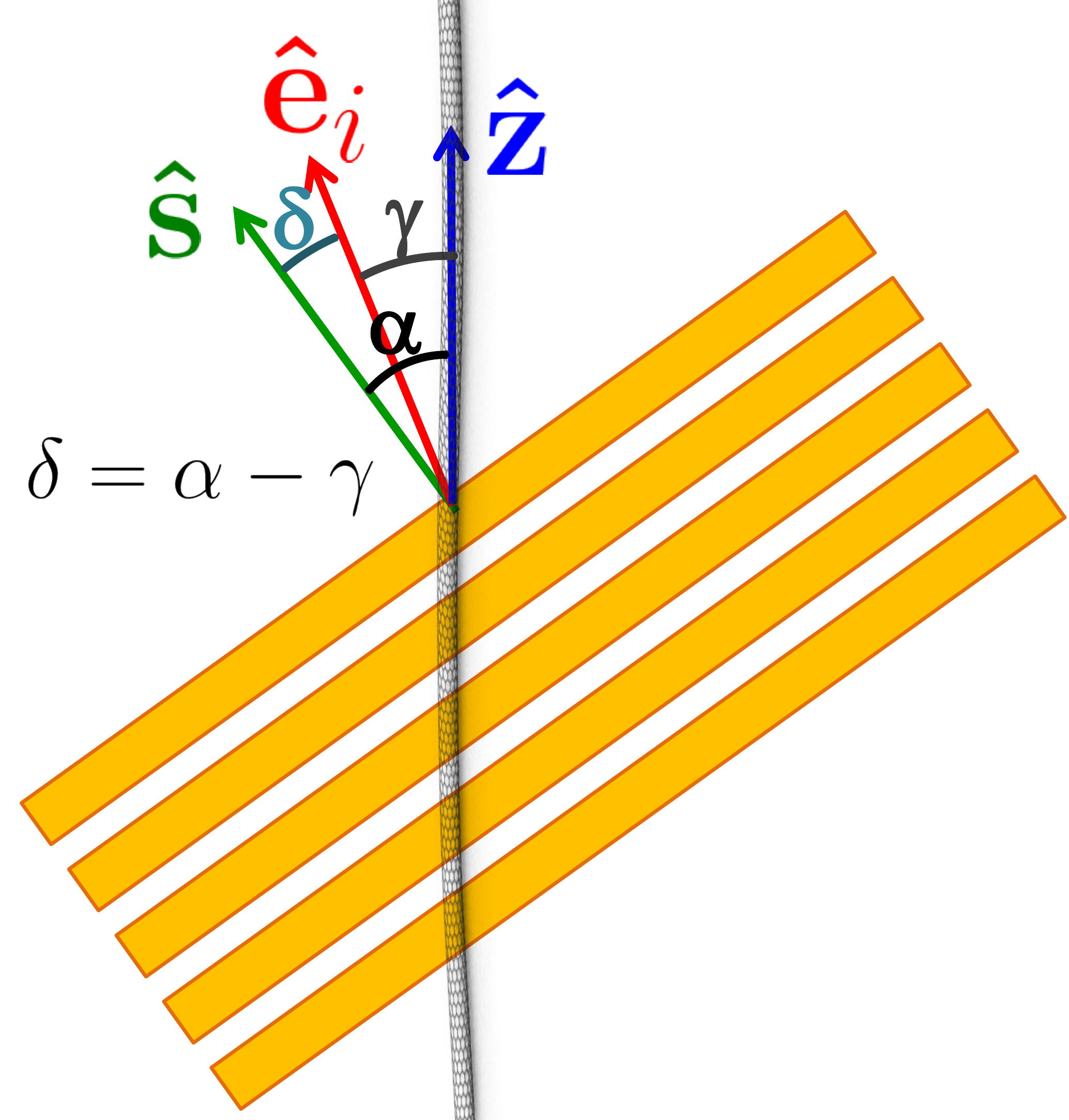}
\caption{When the unit vectors $\mathbf{\hat{s}}$ (array axis), $\mathbf{\hat{z}}$ (CNT axis) and $\mathbf{\hat{e}_i}$ (incident polarization direction) are non-collinear, the Raman process can be approximately described as the result of a series of field projections: the incident electric field $E_0\mathbf{\hat{e}_i}$ is first amplified and projected to $\mathbf{\hat{s}}$ by the antenna array. Then it is projected to $\mathbf{\hat{z}}$ owing to the depolarization effect. At this point, the scattered Raman intensity is then proportional to $\cos^2\gamma\cos^2\alpha$. The Raman scattered light is again amplified and projected to $\mathbf{\hat{s}}$, thus providing a further $\cos^2\alpha$ factor.}
\label{fig:sketchangles}
\end{figure}
In the main text we discuss the impact of the angle $\gamma$ between the CNT axis $\mathbf{\hat{z}}$ and the incident polarization $\mathbf{\hat{e}}_i$, assuming constant the angle $\alpha$ between $\mathbf{\hat{z}}$ and the array axis $\mathbf{\hat{s}}$. In this section we discuss more in detail the general case, when $\mathbf{\hat{s}}$, $\mathbf{\hat{z}}$ and $\mathbf{\hat{e}}_i$ are non-collinear. The situation is sketched in Fig.~\ref{fig:sketchangles}. If the incident light excites a SPP, the field is amplified and rotated toward $\mathbf{\hat{s}}$. The coupling with the antennas (and thus the amplitude of the induced field) is proportional to $\cos\delta$, where $\delta\equiv\alpha-\gamma$. Hence, the intensity of the total field illuminating the antennas is in good approximation proportional to $\cos^2\delta$.

Owing to the depolarization effect, only the component of the resulting field along the CNT axis $\mathbf{\hat{z}}$ will produce a Raman signal, which will then be proportional to $\cos^2\alpha$. The scattered Raman signal is emitted preferentially along the CNT axis. Thus, if also the scattered frequency excites a SPP in the antennas, then this will be again amplified and rotated along $\mathbf{\hat{s}}$. Since only the component of the scattered field along $\mathbf{\hat{s}}$ can excite a SPP, we expect an additional $\cos^2\alpha$ factor. Therefore, under these assumptions, in the general case the measured Raman signal will be proportional to $\cos^2\delta\cos^4\alpha$. In particular for a given $\alpha$, we expect to observe a $\cos^2(\gamma-\alpha)$ dependence, in agreement with the data in Fig.~\ref{fig:pol_dep}.

We stress that it is not easy to quantitatively verify the general $\alpha$-dependence, since this requires to compare different antenna arrays that can be more or less affected by contamination or fabrication imperfections. Moreover, single-wall CNTs are usually not straight and this introduces an uncertainty in the determination of the $\alpha$ angle over the antenna array width. Nevertheless in our sample we observe in most cases a decay of the Raman signal for large $\alpha$ angle,  even though the present uncertainties do not allow us to verify quantitatively the $\cos^4\alpha$ dependence.

\bibliography{biblio}

\end{document}